\documentclass{llncs}
\usepackage{wrapfig}
\usepackage{tikz}
\usetikzlibrary{arrows,backgrounds,positioning}
\usepackage{booktabs}
\usepackage{hyperref}
\usepackage{latexml}
\usepackage[numbers, comma, sort]
{natbib} 
\makeatletter
\renewcommand\bibsection%
{
  \section*{\refname
    \@mkboth{\MakeUppercase{\refname}}{\MakeUppercase{\refname}}}
} 
\makeatother

\usepackage{graphicx}

\hypersetup{
   colorlinks = true,
   urlcolor = blue,
   linkcolor = blue
}

\newcommand{\mlt}[1]{ \begin{tabular}[c]{@{}l@{}}#1\end{tabular}}
\newcommand{\mct}[1]{ \begin{tabular}[c]{@{}c@{}}#1\end{tabular}}

\begin{document}

\title{Growing the Digital Repository of Mathematical Formulae with
Generic \LaTeX\ Sources \thanks{The final publication is 
available at \url{http://link.springer.com}.}}

%
\author{
Howard S. Cohl\,\inst{1}
\and 
   Moritz Schubotz\,\inst{2}
\and
   Marjorie A.~McClain\,\inst{1}
\and
   Bonita V.~Saunders\,\inst{1}
\and 
   Cherry Y.~Zou\,\inst{3}
\and 
   Azeem S.~Mohammed\,\inst{3}
\and 
   Alex A.~Danoff\,\inst{4}
}

\institute{Applied and Computational Mathematics Division,
National Institute of Standards and Technology (NIST),
Gaithersburg, Maryland, U.S.A.,
\email{\{howard.cohl,marjorie.mcclain,bonita.saunders\}@nist.gov}
\and
Database Systems and Information Management Group,
Technische Universit\"{a}t, Berlin, Germany,
\email{schubotz@tu-berlin.de}\\
\and
Poolesville High School, Poolesville, Maryland, U.S.A.,
\email{\{chzou2009,azeemsm\}@gmail.com}
\and
Wootton High School, Rockville, Maryland, U.S.A.
\email{aadanoff@gmail.com}
}
\authorrunning{Cohl, McClain, Saunders, Schubotz, Zou, Danoff, and Mohammed} 

\authorrunning{Cohl, McClain, Saunders, Schubotz, Zou, Danoff and Mohammed}

\clearpage

\maketitle

\vspace{-0.745cm}
\begin{abstract}
One initial goal for the DRMF is to seed our digital compendium with 
fundamental orthogonal polynomial formulae.
We had used the data from the NIST Digital Library of Mathematical 
Functions ({\tt DLMF}) 
as initial seed for our DRMF project. The DLMF input \LaTeX\ source already contains 
some semantic information encoded using a highly customized set of 
semantic \LaTeX\ macros.
Those macros could be converted to content \MathML\ using \LaTeXML.
During that conversion the semantics were translated to an implicit DLMF 
content dictionary.
This year, we have developed a semantic 
enrichment process whose goal is to infer semantic information 
from generic \LaTeX\ sources. The generated context-free semantic 
information is used to build DRMF formula home pages for each 
individual formula.
We demonstrate this process using 
selected chapters from the book ``Hypergeometric Orthogonal 
Polynomials and their $q$-Analogues'' (2010) by Koekoek, Lesky and 
Swarttouw ({\tt KLS}) as well as an actively maintained addendum to this
book by Koornwinder ({\tt KLSadd}).
The generic input {\tt KLS} and {\tt KLSadd} \LaTeX\ sources describe
the printed representation of the formulae, but does not contain 
explicit semantic information.  See \url{http://drmf.wmflabs.org}.
\end{abstract}

\vspace{-1.00cm}
\section{Introduction}
\label{sect:introduction}
\vspace{-0.1cm}

Formula home pages are the principal conceptual objects for the DRMF project.
These should contain the full 
context-free semantic information concerning 
individual orthogonal polynomial and special function (OPSF) formulae.  
The DRMF is designed for a mathematically literate audience and should 
(1) facilitate interaction among a community of mathematicians and scientists 
interested in compendia formulae data for orthogonal polynomials and special 
functions; (2) be expandable, allowing the input of new formulae from the literature;
(3) represent the context-free full semantic information concerning individual 
formulae; (4) have a user friendly, consistent, and hyperlinkable viewpoint 
and authoring perspective; (5) contain easily searchable mathematics; and
(6) take advantage of modern \MathML~tools for easy-to-read, scalably rendered 
content driven mathematics.  In this paper we will discuss the DRMF seeding projects
whose goal is to import data, for example, from traditional print media
(cf.~Figure \ref{fg1}). 

\tikzset{actor/.style={
rectangle,
minimum size=6mm,
very thick,
draw=gray!50!black!50,
top color=white,
bottom color=gray!50!black!20
},
arrow/.style={
-latex, thick, shorten <=2pt,shorten >=2pt
},
darrow/.style={
           <->,
           thick,
           shorten <=2pt,
           shorten >=2pt,
           bend right=45}
}
\begin{figure}[t]
\begin{tikzpicture}[node distance=5mm and 8mm]
\node (Input) [align=center]{plain\\\LaTeX\\ source};
\node (MR) [actor, right=of Input, align=center] {DLMF and DRMF\\macro\\replacement};
\node (SP) [actor, right=of MR, align=center] {Formula\\metadata\\incorporation};
\node (Input2) [align=center, above= of SP]{DLMF \LaTeX\ source};
\node (PKG) [actor, right=of SP, align=center] {Review and\\Packaging \\ Process};
\node (Output) [right=of PKG,align=center]{Wiki\\\XML\\Dump};
\draw[arrow] (Input)--(MR);
\draw[arrow] (MR)--(SP);
\draw[arrow] (Input2)--(SP);
\draw[darrow] (SP)--(PKG);
\path[thick,->] (PKG) edge [bend left] (MR);
\draw[arrow] (PKG)--(Output);
\end{tikzpicture}
\caption{Data flow of seeding projects. For most of the input \LaTeX\ source
distributions, DLMF and DRMF macros are not incorporated.
For the DLMF \LaTeX\ source, the DLMF macros are already incorporated.}
\label{fg1}
\end{figure}
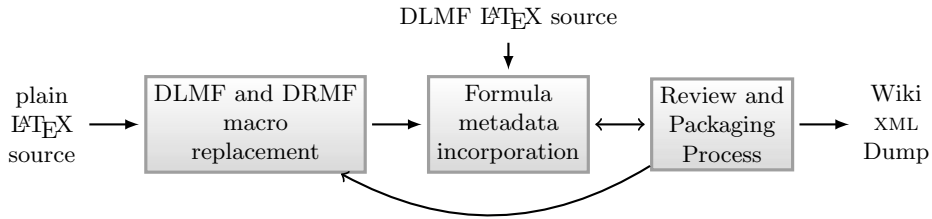

\setcounter{footnote}{0}
We are investigating various sources for seed material in the DRMF 
\cite{CohlDRMF}.
We have been given permission to use a variety of input resources to
generate our online compendium of mathematical formulae.  The current
sources that we are incorporating into the DRMF are given as follows: 
(1) {\it NIST Digital
Library of Mathematical Functions} (\href{http://dlmf.nist.gov}
{{\tt DLMF}}\footnote{We use the typewriter font in this document to refer
to our seeding datasets.})
\cite{NIST,NIST:DLMF}; (2) Chapters 1, 9, and 14 
(a total of 228 pages with about 1800 formulae)
from the Springer-Verlag book {\it ``Hypergeometric
Orthogonal Polynomials and their $q$-Analogues'' (2010) by Koekoek, Lesky and 
Swarttouw} (KLS) \cite{Koekoeketal}; (3) Tom Koornwinder's {\it Additions 
to the formula lists in ``Hypergeometric orthogonal polynomials and their 
$q$-Analogues'' by Koekoek, Lesky and Swarttouw} ({\tt KLSadd}) \cite{KoornwinderKLSadd}; 
(4) Wolfram Computational Knowledge of Continued Fractions Project
(\href{http://blog.wolframalpha.com/2013/05/16/computational-knowledge-of-continued-fractions/}{\tt eCF}); and the {\it Bateman Manuscript Project} ({\tt BMP})
\cite{ErdelyiHTF,ErdelyiTIT} (see Table \ref{tb1}). 
Note that the {\tt DLMF}, {\tt KLS}, {\tt KLSadd}, and {\tt eCF} datasets are currently
being processed within our pipeline.  For the {\tt BMP} dataset, we have furnished high-quality
print scans to Alan Sexton and are currently waiting on the math OCR generated \LaTeX\ output for this
dataset which is currently being generated.
In this paper we focus on DRMF seeding of generic \LaTeX\ sources, namely those which do not 
contain explicit semantic information.

\begin{table}[t]
\centering
\caption{Overview of the first three stages of the DRMF project. Note that the numbers 
which are given are rough estimates.}
\label{tb1}
\begin{tabular}{@{}lccccc@{}}
\toprule
   \hspace{0.9cm}                                    & {\sc Stage} 1  & \hspace{0.9cm}                                            & {\sc Stage} 2                       &\ \ \ \ \ \       & {\sc Stage} 3                                           \\ \midrule
{\sc Started in}                             & 2013                                             & & 2014                                & & ~2015                                             \\ \midrule
{\sc Dataset }                              & \mct{{\tt DLMF}, \\semantic \LaTeX}                 & & \mct{{\tt KLS},\\plain \LaTeX}                 & & \mct{{\tt eCF}: Mathematica\\
{\tt BMP}: book images}
\\ \midrule
\mlt{\sc Semantic \\{\sc enrichment}}                    & \mct{identify constraints,\\substitutions,\\notes, names,\\ proofs, \dots}                &  & \mct{add \\new \\semantic\\ macros}               & & \mlt{image recognition\\ macro suggestion}                                  \\ \midrule
{\sc Technologies}                           & \mct{manual review, \\rule-based \\approaches}     & & improved rules                     & & \mct{natural language
\\processing and  \\machine learning} \\ \midrule
\mlt{ \sc{Number of} \\{\sc formula} \\{\sc home pages}}           & 500                                              & & 1500                                & & 5000                                              \\ \midrule
\mlt{\sc Human time \ \ \ \ \ \ \   \\{\sc per formula} \\{\sc homepage}} & 10 minutes                                       & & 5 minutes                           & & 1 minute                                          \\ \midrule
\mct{\sc Test corpora \\ {\sc contribution}}              & \mct{gold standard \\for constraint \\and proof \\detection}
& & \mct{gold standard \\for \\macro \\replacement} & &        \mlt{evaluation \\ metrics}                                           \\ \bottomrule
\end{tabular}
\end{table}

\section{Seeding with Generic \LaTeX\ Sources}

DRMF seeding projects collect and stream OPSF 
mathematical formulae into formula pages.
Formula pages are classified into those which
list formulae in a broad category, and the individual formula home pages
for each formula.
Generated formula home pages are required to contain bibliographic information 
and usually contain a list of symbols, substitutions and constraints required by the 
formulae, proofs and formula names if available, as well as related notes.  
Every semantic formula entity (e.g., function, polynomial, sequence, operator, 
constant or set) has a unique name and a link to its definition or 
description.  

For \LaTeX\ sources which are extracted from the DLMF project, the semantic macros are 
already incorporated \cite{MillerYoussef2003}.  However, for generic sources such as the 
{\tt KLS} dataset, the semantic macros need to be inserted in replacement for the 
\LaTeX\ source which represents that mathematical object.

Here we give representative examples for the trigonometric sine function, 
gamma function, Jacobi polynomial and little $q$-Laguerre/Wall polynomials, which are rendered respectively
as $\sin z$, $\Gamma(z)$, $P_n^{(\alpha,\beta)}(x)$, and $p_n(x;a|q)$.  These functions and orthogonal
polynomials have \LaTeX\ presentations given respectively by 
\texttt{\textbackslash sin z},
\texttt{\textbackslash Gamma(z)},
\texttt{P\textunderscore n\textasciicircum\{(\textbackslash alpha,\textbackslash beta)\}(x)},
and 
\texttt{p\textunderscore n(x;a|q)}.  The semantic representations for these functions and 
orthogonal polynomials are given
respectively by 
\texttt{\textbackslash sin@@\{z\}},
\texttt{\textbackslash EulerGamma@\{z\}},
\texttt{\textbackslash Jacobi\{\textbackslash alpha\}\{\textbackslash beta\} \{n\}@\{x\}},
\texttt{\textbackslash littleqLaguerre\{n\}@\{x\}\{a\}\{q\}}.
The arguments before the @ or @@ symbols are parameters and the arguments after the @ or @@ 
symbol are in the domain of the functions and orthogonal polynomials. The different between
the @ or @@ symbols indicates a specified difference in presentation, such as the inclusion
of the parentheses or not in our trigonometric sine example.
For the little $q$-Laguerre polynomials, one has three arguments within parentheses.  These
three arguments are separated by a semi-colon and a vertical bar.  Our macro replacement
algorithm indentifies these polynomials, and then extracts the information about what the
contents of each argument is.  Furthermore there are many ways in \LaTeX\ to represent 
open and close parenthese, our algorithm identifies these.  Also, since the vertical bar 
in \LaTeX\ can be represented by `$|$' or `\textbackslash mid', we search for both of these
patterns.  Our algorithm, for instance, also searches for and 
removes all \LaTeX\ white-space characters 
such as those given by \textbackslash, \textbackslash! or \textbackslash hspace\{\}.
There are many other details about making our search and replace work, which we will 
not mention here.

\vspace{-0.2cm}
\section{{\tt KLS} Seeding Project}

In this section we describe how we augment the input {\tt KLS} 
\LaTeX\ source in order to generate formula pages (see Figure \ref{fg1}).
We are developing software processes input \LaTeX\ source to 
generate output \LaTeX\ source with semantic mathematical macros incorporated.
The semantic \LaTeX\ macros that we are using (664 total with 147 currently being 
used for the DRMF project) are being developed by NIST for use in the {\tt DLMF} and 
DRMF projects.  Whenever possible, we use the standardized definitions from 
the NIST Digital Library of Mathematical Functions \cite{NIST}.  If the 
definitions are not available on the {\tt DLMF} website, then we link to 
definition pages in the DRMF with included symbols lists.
One main goal of this seeding project is to incorporate mathematical semantic 
information directly into the \LaTeX\ source.  The advantage of incorporating
this information directly into the \LaTeX\ source is that mathematicians are
capable of editing \LaTeX\ whereas human editing of \MathML~is not feasible.
This enriched information can be further modified by mathematicians using their
regular working environment.

For the 3 chapters of the {\tt KLS} dataset plus the {\tt KLSadd} dataset, a total number 
of 89 semantic macros were replaced a total of 3308 times.
That's an average of $1.84$ macros replaced per formula.  Note that the {\tt KLSadd} dataset
is actively being maintained, and when a new version of it is published, in an automated
fashion, incorporate this new information into the DRMF. This fraction will increase
when more algebraic substitution formulae are included as formula metadata.
The most common macro replacements are given as follows.
The macro for the cosine function, Racah polynomial, Pochhammer symbol, $q$-hypergeometric function,
Euler gamma function, and $q$-Pochhammer symbol were converted a total number of times
equal to 117, 205, 237, 266, and 659. Our current conversions, which use a rule 
based approach, can be quite complicated due to the nature of the variety
of combinations of \LaTeX\ input for various OPSF objects.  
In \LaTeX\ there are many ways of representing parentheses which are usually 
used for function arguments. Also, there are many ways to represent spacing 
delimiters which can mostly be ignored as far as representing the common 
semantic information for a mathematical function call.  Our software canonicalizes
these additional meaningless degrees of freedom and generates easy-to-read semantic
\LaTeX\ source and improves the rendering.  
Developing automatic software which performs macro replacements for OPSF functions 
in \LaTeX\ is a challenging task.  The current status of our rule-based approach is 
highly tailored to our specific {\tt KLS} and {\tt KLSadd} input \LaTeX\ source.

Historically, the desired need for formal consistency has driven mathematicians 
to adopt consistent and unique notations \cite{AAR}. This is extremely beneficial in the 
long run.  We have interacted on a regular basis with the authors of 
the {\tt KLS} and {\tt KLSadd} datasets. They agree that our assumptions about 
consistent notations are correct and they consider using our semantic 
\LaTeX\ macros in future volumes.  Certainly the benefit of using these
macros in communicating with different computer systems is clear.

Once semantic macros are incorporated, the next task is to identify formula
metadata.  
Formula metadata can be identified within and must be 
associated with formulae. One must then identify semantic information 
for the formula within the surrounding text to produce formula 
annotations which describe this semantic information. 
There are annotations which can be summarized as constraints, 
substitutions, proofs and formula names if available, as well as related notes.  
The automated extraction of formula metadata is a challenging aspect 
of the seeding project and future computer implementations might use 
machine learning methods to achieve this goal. However, we have built 
automated algorithms to extract formula metadata.  We have for instance 
identified substitutions by associating definitions for algebraic or 
OPSF functions which are utilized in surrounding formulae.  The automation
process continues by merging these substitution formulae as annotations in
the original formulae which use them.  Another extraction algorithm we have 
developed is the identification of related variables, understanding their 
dependencies and merging corresponding annotations with the pre-existing formula 
metadata. We have manually reviewed the printed mathematics to identify formula
metadata. After we have exhausted our current rule-based approach for extracting
the formula annotations, we will perform the manual insertion of the
missing identified annotations into the \LaTeX\ source.  This will then be followed
by careful checking and expert editorial review.  This also evaluates the 
quality of our rule-based approach and creates a gold standard for future programs.

Once the formula metadata has been completely extracted from the text, then the
remainder of the text should be removed and one is left with a list of \LaTeX\
formulae with associated metadata.  From this list (at the current stage of 
our project), we use this semantic \LaTeX\ source to generate Wikitext.  
One of the features of the generated Wikitext
is that we use a glossary that we have developed of our {\tt DLMF} and DRMF macros
to identify semantic macros within a formula and its associated metadata.
Presentation and meaningful content \MathML~is generated from the {\tt DLMF} 
and DRMF macros using a customized \href{http://dlmf.nist.gov/LaTeXML}{\LaTeXML}
server ({\tt http://gw125.iu.xsede.org}) hosted by the XSEDE project that includes
all generated semantic macros.
From this glossary, we generate symbols lists for each formula which uses 
recognized symbols.  The generated Wikitext is converted to the {\sf MediaWiki} 
\XML-Dump format, which is then bulk imported to our wiki instance.
Our DRMF Wiki has been optimized for \MathML-output.
Because we are using {\sf Mathoid} to render mathematical expressions \cite{Schubotzmathoid},
browsers without \MathML-support can display DRMF formulae within {\sf MediaWiki}.
However, some \MathML-related features (such as copying
parts of the \MathML~output) are not available on these browsers.

At the moment, There are 1282 {\tt KLS} and {\tt KLSadd} wikitext pages.
The current number of {\tt KLS} and {\tt KLSadd} formula home pages is 1219
and the percentage of non-empty symbols lists in formula home pages is given by 98.6 percent.
This number will increase as we continue to merge substitution
formulae into associated metadata and as we continue to expand our macro replacement effort.  
We have detected 208 substitutions which originally appeared as formulae.
We inserted these in an automated fashion into 515 formulae.
The goal of our learning is to obtain a mostly unambiguous content 
representation of the mathematical OPSF formulae which we use.

\vspace{-0.2cm}
\section{Future outlook}

The next seeding projects which we will focus on are those which correspond to 
image and Mathematica inputs (see Table \ref{tb1}).
We have been given permission from Caltech to use the {\tt BMP} dataset within
the DRMF.  In the {\tt BMP} dataset, the original source for data are printed pages of books.
We are currently collaborating on the development of mathematical optical 
character recognition (OCR) software \cite{Sexton} for use in this project.
We plan to utilize this math OCR software to generate \LaTeX\ output which
will be incorporated with the {\tt DLMF} and DRMF semantic macros using our 
developed macro replacement software.  

We are already developing for our next source, namely the incorporation of
the Wolfram {\tt eCF} dataset into the DRMF.  
We have been furnished the {\sf Mathematica} source (also known as 
{\sf Wolfram language}) for this dataset and we are currently developing 
software which translates in both directions from the {\sf Wolfram language} 
to our semantic \LaTeX\ source with DRMF and {\tt DLMF} macros 
incorporated (cf.~Table \ref{tb1}). 

For the {\tt DLMF} source, due to the hard efforts of the
{\tt DLMF} team for more than the past ten years, we already have semantic macros 
implemented, and all that remains is to extract the metadata from the 
text associated with formulae, removing the text after the content 
has been transferred, converting formulae information in tables to lists
of distinct formulae, and generating formula home pages.  We already have mostly
achieved this for {\tt DLMF} Chapter 25 on the Riemann Zeta function and are currently
at work on Chapters 5 (gamma function), 15 (hypergeometric function), 
16 (generalized hypergeometric functions), 17 ($q$-hypergeometric and related
functions) and 18 (orthogonal polynomials) which will ultimately be merged
with the {\tt KLS} and {\tt KLSadd} datasets. Then we will continue to the
remainder of the DLMF chapters.

Once semantic information has been inserted 
into the \LaTeX\ source, there is a huge number of possibilities on how this 
information can be used.  Given that our datasets are collections of OPSF formulae,
we plan on taking advantage of the incorporated semantic
information as an exploratory tool for symbolic and numerical experiments. 
For instance, one may use this semantic content to translate to computer
algebra system (CAS) computer languages such as those used by {\sf Mathematica}, 
{\sf Maple} or {\sf Sage}.  One could then use the translated formulae 
while taking advantage of any of the features available in those 
software packages.  
We should also mention that the DRMF seeding projects generate real 
content \MathML. This has been a huge problem for Mathematics Information Retrieval
research for many years \cite{Nghiemetal2014,KohlhaseSucan2006}.
One major
contribution of the DRMF seeding projects is that they offer quite reasonable
content \MathML.

From a methodological point of view, we are going to develop evaluation metrics 
that measure the degree of semantic formula enrichment. These should be able to 
evaluate new approaches such as mathematical language 
processing \cite{PagelSchubotz} and/or machine learning approaches based 
on the created gold standard.  Additionally, we are considering 
the use of {\sf sTeX} \cite{sTeX}, in order to simplify the definition of 
new macros. Eventually, we can also develop a heuristic which suggests 
new semantic macros based on statistical analysis.

\vspace{0.4cm}
\setcounter{footnote}{1}
\noindent {\bf\large Acknowledgements\,}\footnote{The mention of specific products, trademarks, or brand 
names is for purposes of identification only. Such mention is not to be interpreted in any way 
as an endorsement or certification of such products or brands by the National Institute of 
Standards and Technology, nor does it imply that the products so identified are necessarily 
the best available for the purpose. All trademarks mentioned herein belong to their 
respective owners.}\\[0.4cm]
\noindent We are indebted to Wikimedia Labs, the XSEDE project, Springer-Verlag, 
the California Institute of Technology, and Wolfram Research Inc. for their 
contributions and continued support.  We would also like to thank Roelof Koekoek, Tom 
Koornwinder, Roberto Costas-Santos, Eric Weisstein, Dan Lozier, Alan Sexton, Bruce Miller,
Abdou Youssef, Charles Clark, Volker Markl, George Andrews, Mourad Ismail, and Dmitry 
Karp for their advice, invaluable assistance, and support.

\vspace{-0.2cm}
\label{sect:bib}

\def\cprime{$'$}

\vspace{0.2cm}
\begingroup
\let\clearpage\relax

\endgroup

\end{document}